\documentclass[aps,prl,twocolumn,superscriptaddress,preprintnumbers,amsmath,amssymb,floatfix,10pt]{revtex4-2}
\usepackage{graphicx}
\usepackage{url}
\usepackage[hidelinks]{hyperref}
\usepackage[capitalise]{cleveref}

\newcommand{\pT}{p_\mathrm{T}}
\newcommand{\mll}{m_{\ell\ell}}
\newcommand{\effN}{N_\mathrm{eff}}
\newcommand{\phistar}{\phi^{*}_{\eta}}
\newcommand{\NkLO}{\mathrm{N}^{k}\mathrm{LO}}
\newcommand{\alphas}{\alpha_{\mathrm{s}}}

\begin{document}

\preprint{IFJPAN-IV-2026-23, MCNET-26-35}

\title{Positive-weight moment matching at NNLO and beyond}

\author{B.~Assi}
\email{bassi@perimeterinstitute.ca}
\affiliation{Perimeter Institute for Theoretical Physics, Waterloo,
Canada}
\affiliation{Department of Physics, University of Toronto and Vector Institute, Toronto,
Canada}
\author{R.~Poncelet}
\email{rene.poncelet@ifj.edu.pl}
\affiliation{The Henryk Niewodnicza\'nski Institute of Nuclear Physics,
Krak\'{o}w, Poland}
\author{J.~Whitehead}
\email{james.whitehead@uj.edu.pl}
\affiliation{Jagiellonian University,
Krak\'{o}w, Poland}

\date{\today}

\begin{abstract}
Parton-shower simulations at state-of-the-art perturbative accuracy carry many negative event weights,
which dilute their statistical power.
For many purposes, this makes them unusable.
We formulate a method for the generation of positive-weighted event samples
which reproduce, up to a controllable approximation,
suites of IR-safe observables with
the highest perturbative accuracy of any available calculation.
This method is arbitrarily differential in phase space,
and free from unphysical artefacts such as binning-dependence.
Applied to a leading-order Pythia sample, the method
gives simultaneous NNLO accuracy for
Drell--Yan and $Z+\text{jet}$ observables
in a positive-weighted event sample, for the first time.
The method generalises immediately to higher orders of perturbation theory,
higher multiplicities, and other processes, observables, and event generators.
\end{abstract}

\maketitle

The High-Luminosity LHC programme requires modelling uncertainties
that match its anticipated experimental precision,
and large simulated samples for detector simulation and unfolding.
At present, event samples with the best calculable perturbative accuracy
are computationally prohibitive to process at scale, because of their large fraction of negative event weights.
Negative weights contribute irreducibly to the variance,
diluting the statistical power
of a given simulation budget, potentially catastrophically \cite{HSFGenWG:2020}.

Efforts to address this bottleneck include
mitigation strategies within established NLO matching methods,
MC@NLO \cite{Frixione:2002ik} (\cite{Danziger:2021xvr,Frederix:2020trv,Frederix:2023hom,ElFarkh:2026wtx})
and POWHEG \cite{Nason:2004rx,Frixione:2007vw} (\cite{Nason:2007vt,Alioli:2010qp}),
alternative matching methods
\cite{Jadach:2015mza,Sarmah:2024hdk,vanBeekveld:2025lpz,Nason:2021xke,Che:2026mep},
and post-hoc methods to repair or remove negative weights within event samples
\cite{Nachman:2020fff,Andersen:2020sjs,Andersen:2021mvw,Andersen:2023cku,Andersen:2026ppe,Doherty:2026fli,Palmer:2025jmb,Shyamsundar:2025nzn}.
Despite these innovations,
no existing method combines
state-of-the-art perturbative accuracy with strictly positive event-weights
in a single hadronised sample.
Existing NNLO matching schemes~\cite{Monni:2019whf,Monni:2020nks,Hamilton:2013fea,Karlberg:2014qua,Alioli:2013hqa,Hoche:2014uhw} all carry negative weights.

For processes with resolved jets in the final state,
the `NNLO revolution' of two-loop calculations
and local subtraction algorithms
\cite{Gehrmann-DeRidder:2005btv,Daleo:2006xa,Currie:2013vh,Czakon:2010td,Czakon:2014oma}
implemented in parton-level Monte Carlo codes
\cite{NNLOJET:2024,Czakon:2019tmo}
now enables
state-of-the-art precision calculations with up to two additional final-state emissions
modelled at lower perturbative orders.
In parallel, the `NLL revolution' has led to parton showers which achieve unprecedented formal accuracy
in their all-orders resummation \cite{Dasgupta:2020fwr,Forshaw:2020wrq,Herren:2022jej,vanBeekveld:2022ukn,Assi:2023rbu,Hoche:2024dee,Preuss:2024vyu,Holguin:2026trd}.
For collider phenomenology, these cannot yet be combined.

The characterisation of positive probability measures
from their complete sequence of polynomial moments is a classical problem 
\cite{Haviland1936,Schmudgen2017}.
On a compact domain, a measure is uniquely determined by its complete
collection of polynomial moments, known as the Hausdorff moment problem%
~\cite{Hausdorff1921,HildebrandtSchoenberg1933}.
With finitely many moment constraints, the problem is generally underdetermined.
Among all reweightings of an input sample that reproduce those moments,
the principle of maximum entropy~\cite{Jaynes:1957zza} selects the unique
event-by-event differential reweighting that achieves this
while changing the sample the least.
Provided that the system of constraints is consistent and can be solved,
any solution is positive by construction,
and so preserves the positivity of any positive-definite input sample.

Maximum-entropy reweighting of parton shower samples (`priors') to reproduce precision moments, with
strictly positive weights, was introduced for $e^+e^-$ event shapes~\cite{Assi:2025ibi,Assi:2026tfh}
to augment the leading-logarithmic accuracy achieved by the shower algorithm
to reproduce the state-of-the-art NNLL' resummation available analytically, and has since been applied to resummed Drell--Yan spectra~\cite{Assi:2026tqz}.

We reformulate this approach with constraints from
parton-level Monte Carlo codes.
Because the constraints are calculated independently of their application to showered, hadronised
event samples,
there is no requirement to reformulate NNLO subtraction calculations in a manifestly
positive way, nor to interface NNLO codes to parton showers directly.
As well as new runs of NNLO codes, moments can be computed
from cached event samples \cite{Bern:2013zja,Czakon:2023hls},
or approximately from the interpolation grids used for PDF fitting
\cite{Carli:2010rw,Carrazza:2020gss,Britzger:2022lbf,Jezo:2026adf}.
There is no material limit to the number of observables,
perturbative order,
multiplicity
or number of higher moments that can be incorporated,
beyond their availability to within the requisite statistical uncertainty.
We return to this point below.

We emphasise that the reweighting process is \textit{systematically improvable}.
When further multiplicities, observables, or moments become known
to higher perturbative accuracies or statistical precision,
existing samples can be refined quickly and cheaply to match.
This is a paradigm shift from the current approach, in which the accuracy of
large samples is locked-in when they are generated.

We demonstrate the method using an LO+PS Pythia Drell--Yan event sample,
achieving
simultaneous moment-matched NNLO accuracy for inclusive observables
and for $Z+\text{jet}$ observables.
Rather than saturate the dimensionality of the fixed-order phase space,
we match only a subset of the observables of interest, in order to
validate the method on held-out observables not used in the reweighting.
We benchmark the computational cost and statistical power of the resulting sample
against alternative methods.

\emph{Method.}
Let the prior $\{\Phi_i,p_i\}$ consist of events $\Phi_i$ with
weights $p_i>0$, $\sum_i p_i=1$, and let $\boldsymbol{m}(\Phi)$ be a vector of
measurement functions, $\boldsymbol{m}_i\equiv\boldsymbol{m}(\Phi_i)$, whose
fixed-order expectations $\boldsymbol{\mu}$ are calculated to $\NkLO$.
Among all normalized weights
$q_i\geqslant0$ that reproduce $\boldsymbol{\mu}$, we take the one closest to the
prior in relative entropy~\cite{Shore:1980,Csiszar:1975},
\begin{equation}
\min_{q}\,\sum_i q_i\ln\frac{q_i}{p_i}
\ \ \text{with}\ \
\sum_i q_i\,\boldsymbol{m}_i=\boldsymbol{\mu},\ \sum_i q_i=1.
\label{eq:kl}
\end{equation}
The objective is strictly convex, so a solution, where one exists, is unique,
and has the closed form
\begin{equation}
q_i = \frac{p_i\, e^{\boldsymbol{\lambda}\cdot\boldsymbol{m}_i}}{\mathcal{Z}(\boldsymbol{\lambda})},
\qquad
\mathcal{Z}(\boldsymbol{\lambda}) = \sum_j p_j\, e^{\boldsymbol{\lambda}\cdot\boldsymbol{m}_j} ,
\label{eq:maxent}
\end{equation}
i.e.\ is manifestly positive for every multiplier vector $\boldsymbol{\lambda}$.
We call $q_i/p_i$ the \emph{tilt}.
The multipliers follow from a strictly convex dual
problem, which may be solved cheaply by Newton iteration (End Matter).
As for any multiplicative reweighting, the prior must have support
throughout the target region, or the multipliers diverge.
For the Drell--Yan results reported here, the tilts $q_i/p_i$ lie
between 0.97 and 1.79 for the NLO $Z\!+\!j$ input and
between 0.96 and 3.95 for the NNLO one.

As constraints we use Chebyshev polynomial moments
of observables $v(\Phi)$,
mapped monotonically to $[-1,1]$ by $u_v$, calculated event-by-event.
The map $u_v$ is chosen linear for rapidity variables,
logarithmic for steep spectra,
and the Breit--Wigner coordinate for resonances (End Matter).
A window $w_v$ restricts each input calculation to its region of theoretical accuracy.
Each constrained observable contributes a \emph{tower} of
moments, whose targets blend the prior outside the window with the fixed order inside it,
\begin{align}
\boldsymbol{m}&\supset\big\{T_n(u_v)\big\}_{n=1}^{N_v},
\nonumber\\
{\mu_a}
&
{{}=(1-r_v)\,\langle m_a\rangle_{p,\,1-w_v}+r_v\,\langle m_a\rangle_{\mathrm{FO},\,w_v}},
\label{eq:phigen}
\end{align}
where $T_n$ are the Chebyshev polynomials,
$a$ indexes the moments,
$r_v$ is the fixed-order share of the cross-section inside the window,
and the averages are weighted by $1-w_v$ over the prior,
and $w_v$ over the fixed-order moments.
Only expectations in the span of the imposed measurement functions are fixed directly.
Other predictions depend on the finite-moment approximation and on conditional distributions inherited from the prior.
Where $w_v$ vanishes the target is the prior's own moment, where it is one the fixed-order moment.
Since
\begin{equation}
	\ln\left( \frac{q_i}{p_i} \right)
	=
	\sum_v P_v(u_v) - \ln\mathcal{Z}
\end{equation}
with $P_v$ a polynomial of
degree $N_v$, the tilt is smooth by construction, and depends
only on the constrained observables.
For fixed values of these observables, the conditional 
distribution of the remaining event degrees of freedom
remains that of the prior.
Where a correlation is itself a perturbative target,
observables can be constrained jointly through cross-moments (End Matter).

\emph{Uncertainty.}
If moments are assumed to be known exactly (i.e., taking the large-statistics limit),
the convergence of the reweighted distributions to the target
is governed by how well polynomials can approximate the log-ratio of target to prior densities~\cite{doi:10.1137/0801014}.
Soft-gluon radiation, as generated by a parton shower,
smooths the local discontinuities of fixed-order calculations~\cite{Catani:1997xc}.
Provided that kinematic thresholds are mapped to the endpoint of the Chebyshev domain,
the target and prior are both smooth and strictly positive,
and thus the logarithm of their ratio is real and smooth.
This is the function approximated by MaxEnt as a polynomial,
and so the reweighted distribution converges uniformly
to the target as more moments are included,
and under stronger regularity assumptions on the prior, geometrically.
In particular, provided the prior has support everywhere the target does,
the Kullback--Leibler divergence \cite{Kullback:1951zyt} between the target and the MaxEnt reweight with Chebyshev moments truncated at order $N_v$ converges as $\mathcal{O}(\rho^{-2N_v})$ under favourable regularity assumptions.
The total variation $d_{\mathrm{TV}}$ converges as $\mathcal{O}(\rho^{-N_v})$.
Here $\rho>1$ labels the Bernstein ellipse $E_\rho$, in whose neighbourhood the log-ratio of target to prior is analytic (End Matter).

In practice, higher moments are harder to resolve statistically and so
the moment-truncation uncertainty must be balanced against the
statistical and scale uncertainty on the higher moments.
We incorporate moment constraints up to $N_v$ in each tower,
the highest order that is resolved at the level of the target moments' statistical convergence.
Order $n$ is considered resolved when the shift it incurs exceeds
the input's combined Monte Carlo and scale uncertainty $\sigma_\mathrm{FO}$,
i.e., when its signal-to-noise ratio
\begin{equation}
	\label{eq:SNR}
	\mathrm{SNR}_n
	=
	\frac{|\mu_n^\mathrm{FO}-\mu_n^\mathrm{prior}|}{\sigma_\mathrm{FO}(\mu_n)} > 1,
\end{equation}
with $\mu_n$ the tower's $n$\textsuperscript{th} moment normalised
to its window, on the fixed-order points and on the prior.

\begin{figure}[!t]
\includegraphics[width=\columnwidth]{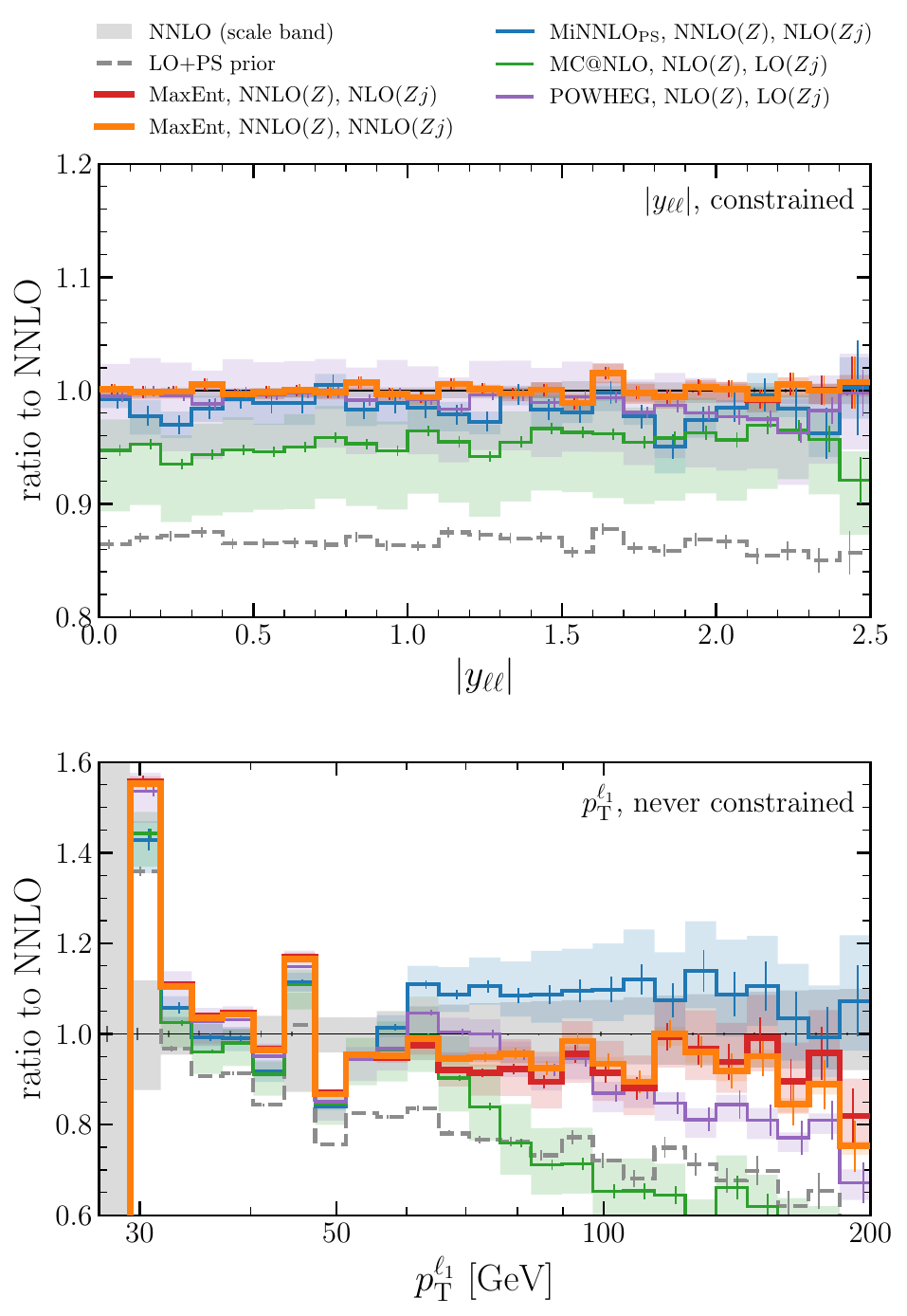}
\caption{Ratios to NNLO of the differential cross-sections of two observables,
	rapidity $|y_{\ell\ell}|$ (constrained, top) and leading-lepton transverse momentum
	$\pT^{\ell_1}$ (unconstrained, bottom).
	The reweighted LO+PS sample (MaxEnt, thick red and orange)
	reproduces the input NNLO distribution to the sub-percent level.
	On the held-out observable $\pT^{\ell_1}$  the reweighting procedure
	corrects the prior to good agreement with the NNLO calculation,
	despite the absence of an explicit constraint.
	Here and throughout, bands represent
	seven-point scale-variation envelopes,
    and bars statistical errors (End Matter).}
\label{fig:closure}
\end{figure}

\emph{Numerical results.}
\begin{figure*}[t]
\centering
\includegraphics[width=0.495\linewidth]{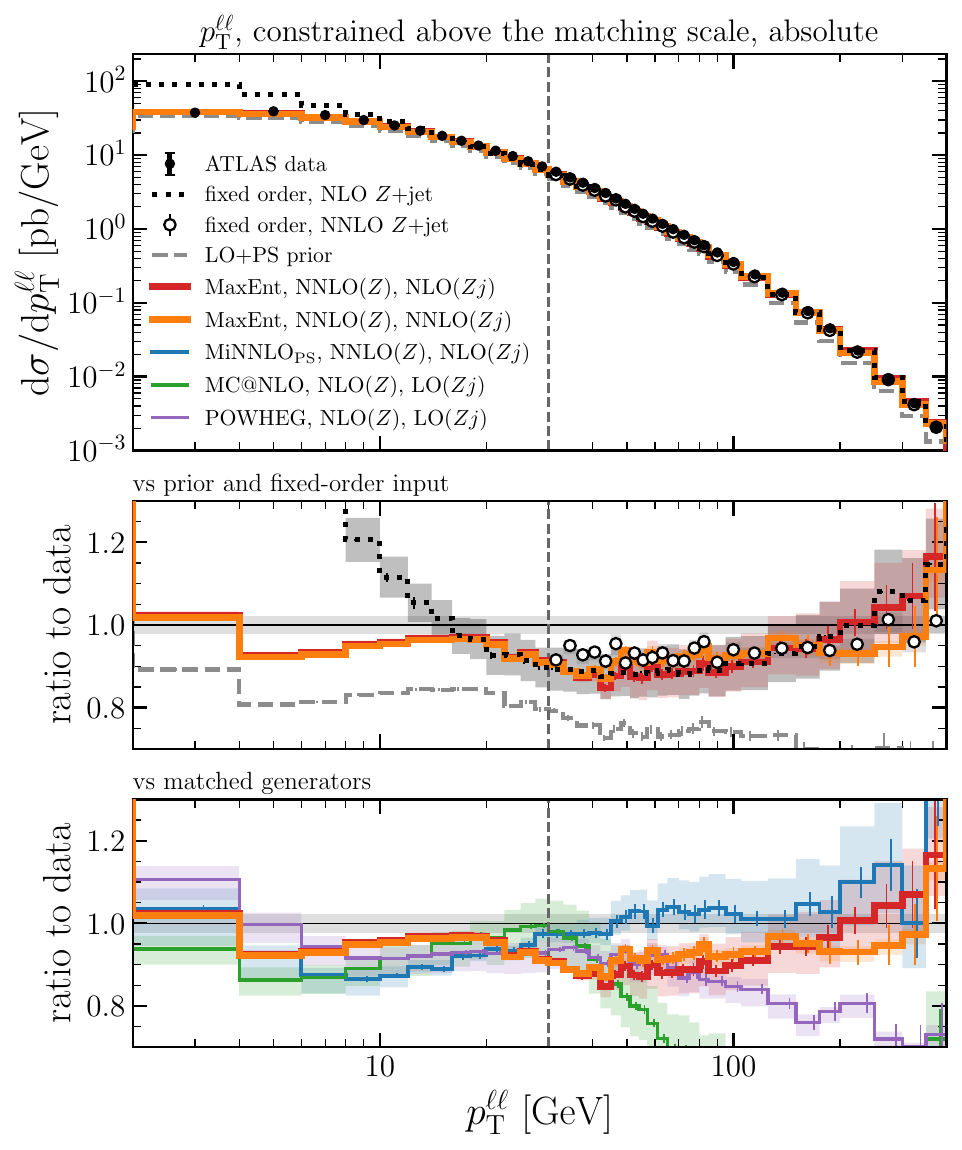}\hfill
\includegraphics[width=0.495\linewidth]{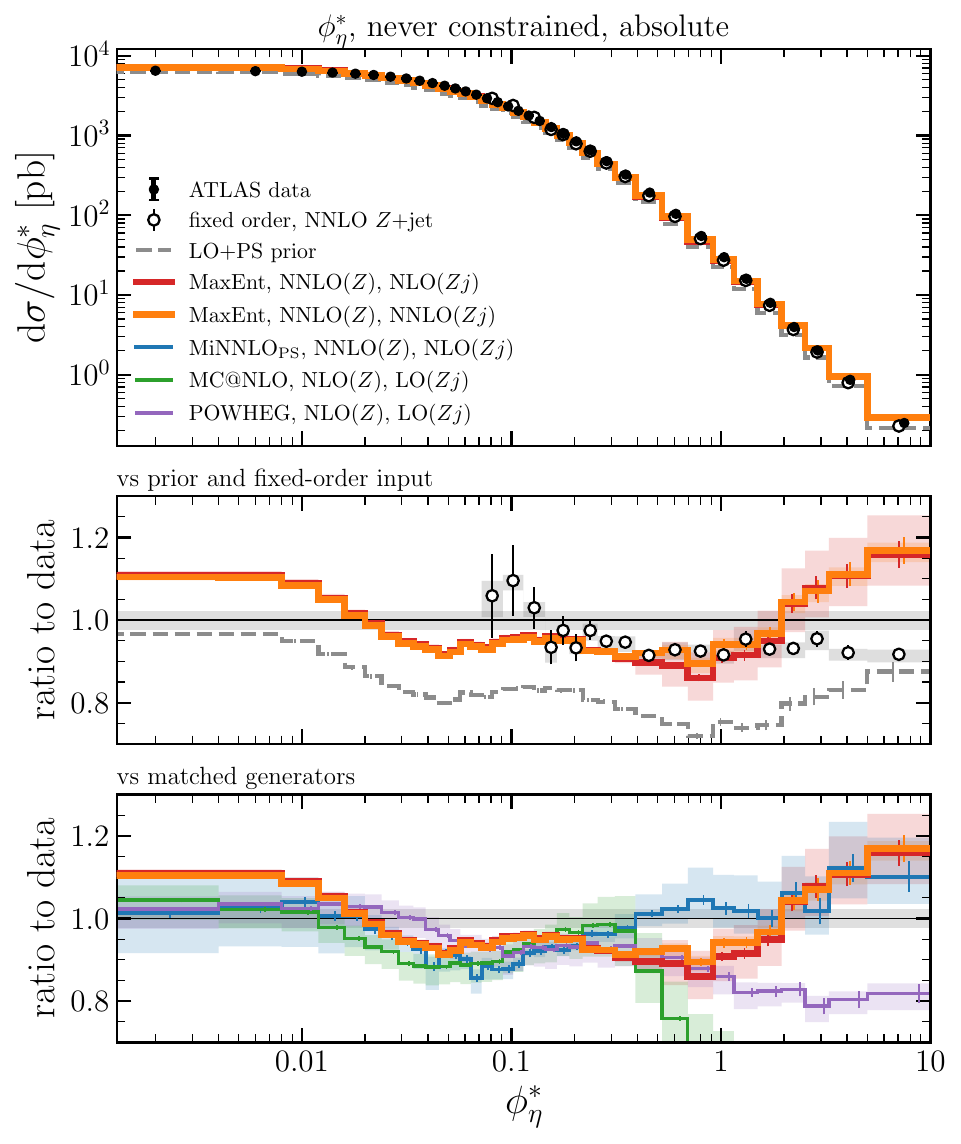}
\caption{MaxEnt, the LO+PS prior, fixed-order NNLO DY and $Z\!+\!j$, NLO-matched MC@NLO and POWHEG, and NNLO-matched MiNNLO$_\mathrm{PS}$ predictions, compared to 13\,TeV ATLAS data~\cite{ATLAS:2019zci},
for the constrained $Z\!+\!j$ observable $\pT^{\ell\ell}$ (left)
and the unconstrained observable $\phistar$ (right).
The legends give the order of each sample for the inclusive process and for the recoil. Cross sections are absolute, the data being the published shape times the measured fiducial cross section, $736\pm17$~pb.
Both MaxEnt variants, with or without the NNLO $Z\!+\!j$ moment constraints,
describe the data well across the kinematic range. They reconcile fixed-order accuracy above the transition window, whose lower edge is the dashed vertical line, with the resummation of the parton shower prior below it.
On the right only the NNLO $Z\!+\!j$ calculation is shown, from $\phistar=0.07$, where its $5\,$GeV cut on $\pT^{\ell\ell}$ leaves it complete. Below $\phistar=0.15$ its bins are merged in pairs.}
\label{fig:money}
\end{figure*}
We use moments calculated with NNLOJET~\cite{NNLOJET:2024}
for the inclusive Drell--Yan observables, choosing
$\mu_\mathrm{R}\!=\!\mu_\mathrm{F}\!=\!m_{\ell\ell}$,
and with STRIPPER~\cite{Czakon:2010td,Czakon:2014oma,Czakon:2019tmo}
for the $Z\!+\!j$ observables,
with
$\mu_\mathrm{R}\!=\!\mu_\mathrm{F}\!=\!m_\mathrm{T,\ell\ell}$.
We vary the renormalisation and factorisation scales by the standard seven-point prescription
as a heuristic estimate of missing higher-order uncertainties, and propagate it through the MaxEnt moment-matching.

The prior used here is a sample of $2.1\times10^{6}$ events generated as
$p p \to \ell^+ \ell^-$
at leading-order within \texttt{MG5\_aMC}~\cite{Alwall:2014hca},
showered by Pythia~8.312~\cite{Sjostrand:2022},
with multiparton interactions and hadronisation on and matrix-element corrections off,
of which $7.5 \times 10^5$ pass the ATLAS fiducial cuts~\cite{ATLAS:2019zci}, $\pT^{\ell}>27\,\mathrm{GeV}$, $|\eta^{\ell}|<2.5$ and $66<\mll<116\,\mathrm{GeV}$.
All samples are generated with a mass window only, and the fiducial cuts are applied after the shower, so that the shower recoil can move leptons across the fiducial boundaries.
For the input moments and the hard process of the prior we use
$\sqrt{s} = 13\, \mathrm{TeV}$, the CT18NNLO PDF set~\cite{Hou:2019efy},
and for the input moments we use $\alphas(M_Z) = 0.118$.

As references we use MC@NLO in \texttt{MG5\_aMC}~\cite{Alwall:2014hca} and POWHEG at NLO,
and MiNNLO\textsubscript{PS}~\cite{Monni:2019whf,Monni:2020nks} at NNLO,
the last two in \texttt{POWHEG-BOX-V2}~\cite{Alioli:2010xd},
all with the same PDF set, $G_\mu$-scheme electroweak inputs and no folding.

We compare with the ATLAS 13\,TeV measurement~\cite{ATLAS:2019zci} of the $Z$-boson transverse momentum, precise to about $0.2\%$ below
$30\,\mathrm{GeV}$, and held-out (but kinematically correlated) observable $\phistar$~\cite{Banfi:2010cf}.

For the inclusive observables we apply no window beyond the fiducial cuts,
i.e.\ $w_v\!\equiv\!1$.
For resolved $Z\!+\!j$ observables $w_v$ rises from zero at the matching threshold $v_\mathrm{m}$ to one at $2v_\mathrm{m}$,
and is cut off above $500\,\mathrm{GeV}$ (End Matter).
For $\pT^{\ell\ell}$ we choose $v_\mathrm{m}$ according to
the breakdown of fixed-order perturbation theory in the unresolved $\pT^{\ell\ell} \to 0$ limit,
since the resummation of the parton shower prior provides the more accurate description.
We require $\alphas(v)\ln^{2}(Q/v)<0.2$, which holds above $28\,\mathrm{GeV}$ for
$Q=m_Z$, and accordingly set $v_\mathrm{m}=30\,\mathrm{GeV}$.
Below this threshold the spectrum remains the shower prior's, up to
the effect of the other constraints.

From the NNLO DY moment calculation, the inclusion criterion of \cref{eq:SNR}
selects $12$ moments of $\mll$,
$11$ of $|y_{\ell\ell}|$,
and $20$ moments of $\pT^{\ell\ell}$,
with no joint moments
(Eq.~\eqref{eq:constraints} in the End Matter).
The calculation achieves NNLO accuracy for the inclusive observables, and NLO accuracy for $\pT^{\ell\ell}$
(which vanishes in the DY Born kinematics).
We additionally calculate the $\pT^{\ell\ell}$ moments at NNLO accuracy in a separate NNLO $Z\!+\!j$ calculation.
In this case, the criterion of \cref{eq:SNR} selects 9 moments.

\Cref{fig:closure} illustrates the ratio of the methods to fixed-order NNLO for two inclusive
observables,
$\mathrm{d} \sigma / \mathrm{d} \lvert y_{\ell\ell} \rvert$ (constrained within MaxEnt)
and
$\mathrm{d} \sigma / \mathrm{d} \pT^{\ell_1}$ (unconstrained).
For the moment-constrained observable, the fixed-order distribution is reproduced
to within the propagated statistical uncertainty,
despite the starkly different behaviour of the prior at large rapidities.
From the held-out observable $\pT^{\ell_1}$, both MaxEnt samples can be seen to closely reproduce
the fixed-order distribution except in the low-$\pT^{\ell_1}$ threshold region,
which is resummation-sensitive.
Despite the absence of explicit constraints, the MaxEnt result closely follows the other matched calculations there, and agrees with the fixed-order calculation for larger values of $\pT^{\ell_1}$.
This is in part attributable to the low number of free dimensions in phase space. At Born level $\pT^{\ell_1}$ is fixed by $\mll$ and $|\cos\theta^{*}|$ in the Collins--Soper frame~\cite{Collins:1977iv}, so
the prior's pre-existing correlations are combined with the
additional higher-order information carried by the tilt,
moving $\pT^{\ell_1}$ towards the NNLO prediction.
Additionally imposing the joint moments of $\mll$ and $|\cos\theta^{*}|$ leaves
$\pT^{\ell_1}$ essentially unchanged (End Matter).
In principle this can be approached fully systematically to identify a complete set of
kinematic variables characterising the fixed-order phase-space,
and including higher-order correlation information calculable perturbatively,
such as with Energy-Flow Polynomials (EFPs) \cite{Komiske:2017aww,Assi:2026tfh}.

\Cref{fig:money} similarly shows the distributions
$\mathrm{d} \sigma / \mathrm{d} \pT^{\ell\ell}$ (constrained within MaxEnt,
at NLO and NNLO respectively)
and
$\mathrm{d} \sigma / \mathrm{d} \phistar$
(unconstrained), this time
alongside ATLAS data.
Again, the underlying kinematic correlations between observables present in the prior
propagate perturbative information from $\pT^{\ell\ell}$ into $\phistar$ automatically.
In the last bins, $\phistar\gtrsim2$, the MaxEnt samples and MiNNLO$_\mathrm{PS}$ lie above the data and the NNLO $Z\!+\!j$ prediction.
Above $2v_\mathrm{m}$, where $w_v$ reaches one, the MaxEnt samples follow their fixed-order input within its uncertainty and describe the data as well as the matched generators. Below $v_\mathrm{m}$ they keep the shower's description and avoid the divergence of fixed order in the singular limit.

\emph{Performance comparison.}
\begin{figure*}[tb]
\includegraphics[width=0.95\linewidth]{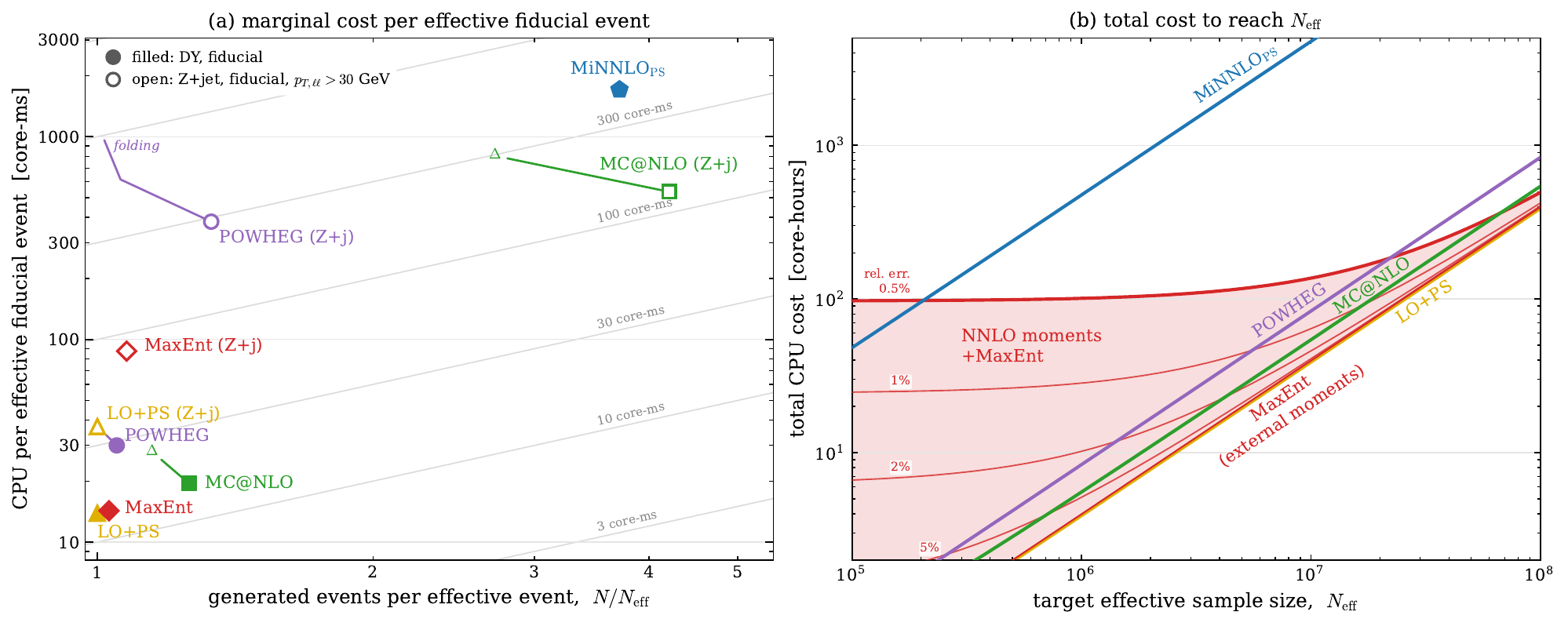}
\caption{
	Performance comparison of alternative generation methods
	(a) per marginal additional effective event, and
	(b) for a generation campaign in which NNLO moments must also be calculated \textit{de novo}.
	The diagonal lines mark constant per-generated-event cost.
	On the left, we additionally show the effect of weight-reduction techniques such as MC@NLO-$\Delta$
	\cite{Frederix:2020trv}
	and folding in POWHEG \cite{Alioli:2010qp}.
	On the right, we show the statistical cost of the NNLO DY campaign with target uncertainties
	down to 0.5\%.}
\label{fig:cost}
\end{figure*}
The variance penalty 
of event samples
of non-uniform weight is given by the Kish effective sample size $\effN$~\cite{Kish:1965},
\begin{equation}
\effN = \frac{\big(\sum_i q_i\big)^{2}}{\sum_i q_i^{2}}
\; \leqslant \;
N \,(1-2f_-)^{2},
\label{eq:neff}
\end{equation}
where $f_-$ is the fractional contribution of negative-weighted events,
\begin{equation}
    f_- = - \frac{\sum_i \Theta[q_i < 0] \; q_i }{\sum_i \lvert q_i \rvert}.
\end{equation}
For weights of equal magnitude $f_-$ is simply the
fraction of negative-weight events, and the bound is saturated.
The tilt keeps $f_- = 0$ by construction, and the reweighting leaves the effective sample size of the NNLO MaxEnt sample essentially unchanged.

To compare the alternative methods on a like-for-like basis we compare them
on a CPU-per-$\effN$ basis,
within the fiducial DY and $Z\!+\!j$ phase spaces used above.
This measures the cost of a statistically-equivalent generation campaign.
We show this comparison in \cref{fig:cost}.
If the moments are known externally, MaxEnt is both computationally cheaper, and delivers
a more statistically-powerful sample, than other methods.
This computational advantage is distinct from
the theoretical advantage of incorporating higher-accuracy perturbative information
for higher-multiplicity observables.
If they must also be calculated \textit{de novo}, this fixed cost is amortised by $\effN\!\sim\!10^8$
relative to existing NLO matching methods, and $10^6$ for NNLO methods.
If an NNLO fixed-order calculation is in any case anticipated, the moment calculation
may be included at negligible additional cost.

\emph{Outlook.}
We have shown that MaxEnt reweighting combines cheap, abundant LO+PS samples
with state-of-the-art perturbative calculations, within a systematically improvable approximation.
The result is a positive-weight sample with higher-order accuracy in suites of distributions,
here NNLO across two multiplicities at once, at negligible additional computational cost.
The method is prior-agnostic.
Because it keeps the correlations of the prior outside the constrained observables,
it can be expected to preserve some of the properties of NLL showers.

Unlike existing methods, MaxEnt can be applied to any process or perturbative order
for which a fixed-order calculation exists,
and samples can be systematically refined to incorporate new or
better-converged results.
Pending the construction of a general fully-differential matching method,
this meets the pressing need for novel methods to address the computational
bottleneck anticipated for Monte Carlo sample generation at the HL-LHC.

\begin{acknowledgments}
\emph{Acknowledgments.}
We thank the organisers of the `Negative-weights suppression in Monte Carlo samples'
workshop held at CERN, at which this work was conceived.
Research at the Perimeter Institute is supported by the Government of
Canada through the Department of Innovation, Science and Industry Canada, and by the
Province of Ontario through the Ministry of Colleges and Universities. 
B.A.\ was supported in part by the Canadian AI Safety Institute Research Program at CIFAR. 
This work was performed in part at the Aspen Center for Physics, with support for B.A.\ by a grant from the Simons Foundation (1161654, Troyer).
R.P.\ is funded by the European Union (ERC, STAPLE, 101222253). Views and opinions expressed are however those of the author(s) only and do not necessarily reflect those of the European Union or the European Research Council. Neither the European Union nor the granting authority can be held responsible for them.
This work was performed in part using the Cambridge Service for Data Driven Discovery (CSD3), part of which is operated by the University of Cambridge Research Computing on behalf of the STFC DiRAC HPC Facility (www.dirac.ac.uk). The DiRAC component of CSD3 was supported by STFC grants ST/P002307/1, ST/R002452/1 and ST/R00689X/1.
J.W. acknowledges the support of grant 2025/57/B/ST2/04034 of the National Science Centre,
Poland
and
the Polish high-performance computing infrastructure PLGrid
(HPC Centre: ACK Cyfronet AGH) for providing facilities and support within computational grant PLG/2026/020231.
\end{acknowledgments}

\emph{Data availability.}
The MaxEnt code and all NNLO moments used here are openly available at
\url{https://github.com/benleo12/maxent_NNLO}.
The event samples are available from the authors upon reasonable request.

\clearpage

\onecolumngrid
\section*{End Matter}
\twocolumngrid

\emph{Solver.}
The multipliers minimize the strictly convex dual
$g(\boldsymbol{\lambda})=\ln\mathcal{Z}(\boldsymbol{\lambda})-\boldsymbol{\mu}\cdot\boldsymbol{\lambda}
+\tfrac{\epsilon}{2}\|\boldsymbol{\lambda}\|^{2}$, with
measurement functions standardized under the prior, by safeguarded Newton
iteration. 
The ridge $\epsilon\!=\!10^{-4}$ makes the Hessian positive definite,
relaxing each moment by $\epsilon\lambda_a$ in prior standard deviations
(significantly below MC errors).
Those errors enter through the depth of each tower
via \cref{eq:SNR}, and the bootstrap re-solves below.

\emph{Maps, window, and targets.}
The Drell--Yan moment constraint set used is
\begin{align}
\boldsymbol{m}={}&\big\{T_n(u_\mathrm{BW}(\mll))\big\}_{n=1}^{12}
\cup\big\{T_n(u_\mathrm{lin}(|y_{\ell\ell}|))\big\}_{n=1}^{11}
\nonumber\\
&\cup
\big\{T_n(u_\mathrm{log}(\pT^{\ell\ell}))\big\}_{n=1}^{20}
\label{eq:constraints}
\end{align}
with 
the recoil targets blended through the window $w$ of Eq.~\eqref{eq:window}, as in Eq.~\eqref{eq:phigen},
and the maps,
for $u_\mathrm{lin}$, $u_\mathrm{log}$, and $u_\mathrm{BW}$ respectively,
\begin{align}
u_v&=2\,\frac{h(v)-h(v_\mathrm{lo})}{h(v_\mathrm{hi})-h(v_\mathrm{lo})}-1,\nonumber\\
h(v)&=v,\ \ \ln v,\ \ \arctan\frac{v^{2}-m_Z^{2}}{m_Z\Gamma_Z},
\label{eq:maps}
\end{align}
on the intervals
$[v_\mathrm{lo},v_\mathrm{hi}]=[66,116]\,\mathrm{GeV}$ for $\mll$, $[0,2.4]$
for $|y_{\ell\ell}|$, and $[30,2500]\,\mathrm{GeV}$ for $\pT^{\ell\ell}$.
A value outside its interval is set to the nearest edge.
In the Breit--Wigner coordinate a pure resonance is flat.
Any monotonic map reaches the same tilt in the limit,
but not under truncation.
The window is the $C^{2}$ profile
\begin{equation}
w(v)=\big(6s^{5}-15s^{4}+10s^{3}\big)\,\Theta(v_\mathrm{c}-v),
\label{eq:window}
\end{equation}
with $s=\min\{1,\max[0,\log_2(v/v_\mathrm{m})]\}$ and $\Theta$ the step
function, rising from $0$ at $v_\mathrm{m}=30\,\mathrm{GeV}$ to $1$ at
$2v_\mathrm{m}$ and closing at $v_\mathrm{c}=500\,\mathrm{GeV}$, above which
the prior's shape is kept.
The threshold $0.2$ in the rule $\alphas(v)\ln^{2}(Q/v)<0.2$ is a scheme choice
that in principle induces an uncertainty.

\emph{Moment depth and truncation.}
Moments are booked to the 12\textsuperscript{th} order for inclusive observables and to the 20\textsuperscript{th} for $p_{\mathrm{T}}^{\ell\ell}$.
We impose all orders up to the highest deemed resolved.
In the NNLO $Z$+jet tower, dropping the unresolved 3\textsuperscript{rd} and 6\textsuperscript{th} moments
moves them by less than their error and changes the recoil above $v_\mathrm{m}$ by less than half a percent in the median bin, well inside the scale band.
We estimate the effect of unknown moments by solving again with two and four fewer orders and taking the change as the truncation uncertainty.
For the inclusive observables it is comparable to the residual deviations, which therefore sit at the noise floor of the input.
For $p_{\mathrm{T}}^{\ell\ell}$, cutting the tower to the nine orders of the NNLO $Z$+jet tower moves the medians of \cref{tab:samples} by less than their errors.

\begin{table}[tb]
\caption{\label{tab:samples}
Negative-weight fraction $f_-$, $\effN/N$, median deviation from the data of the normalised
$\pT^{\ell\ell}$ (below $400\,\mathrm{GeV}$ and above $v_\mathrm{m}$) and
$\phistar$ (all bins and above $0.33$), and the fiducial cross section per lepton flavour,
measured as $736\pm17$~pb.}
\begin{ruledtabular}
\begin{tabular}{lccccccc}
 & $f_-$ & $\effN/N$ & \multicolumn{2}{c}{$\pT^{\ell\ell}$ (\%)} & \multicolumn{2}{c}{$\phistar$ (\%)} & $\sigma_\mathrm{fid}$ \\
 & (\%) & (\%) & ${<}400$ & ${>}v_\mathrm{m}$ & all & ${>}0.33$ & (pb) \\
\hline
LO+PS prior                        & 0    & 100  & 10.5 & 11.8 & 3.4 & 8.3  & 616 \\
MaxEnt, NLO$(Zj)$                  & 0    & 99.9 & 6.0  & 8.0  & 4.0 & 7.7  & 710 \\
MaxEnt, NNLO$(Zj)$\footnote{In six of the nine NNLO $Z$+jet moments the statistical error exceeds the scale half-range, by up to a factor of $1.9$.
}
                                   & 0    & 99.9 & 3.8  & 4.1  & 3.7 & 5.5  & 710 \\
MiNNLO$_\mathrm{PS}$               & 23.5 & 27.2 & 6.9  & 7.7  & 6.1 & 7.8  & 699 \\
MC@NLO                             & 5.2  & 80.4 & 10.5 & 24.2 & 5.1 & 31.7 & 677 \\
POWHEG                             & 1.2  & 95.2 & 4.5  & 8.0  & 4.6 & 13.4 & 705 \\
\end{tabular}
\end{ruledtabular}
\end{table}

\emph{Convergence and moment-truncation uncertainty.}
We sketch the argument underpinning the convergence claim.
Suppose $p$ is the prior probability distribution and $r$ the target.
Let $\ell\!=\!\log (r / p)$ be the exact reweight and suppose that
$\ell_N\!\in\!\mathcal{P}_N$ is the polynomial of degree $N$ or less
that best uniformly approximates $\ell$,
with approximation error $\varepsilon_N = \lVert \ell - \ell_N \rVert_\infty$.
Let $q_N$ be the MaxEnt fit matching all polynomial moments up to degree $N$.
Since $\log(q_N/p) \in \mathcal{P}_N$,
and the moments up to $N$ must match,
\begin{align}
    & 4 d_{\mathrm{TV}}(r, q_N)^2
    \leqslant
    D_\mathrm{KL}(r \Vert q_N)
    + D_\mathrm{KL}(q_N \Vert r) \\ \notag
    & {} = \int (r-q_N) (\ell - \ell_N) 
    \leqslant \varepsilon_N \int \lvert r - q_N \rvert
    = 2 \varepsilon_N d_{\mathrm{TV}}(r, q_N),
\end{align}
applying Pinsker's inequality \cite{Pinsker:1964,Csiszar:1967} for the lower bound.
It follows that the total variation $d_{\mathrm{TV}}(r, q_N) \leqslant \frac{1}{2} \varepsilon_N$ and
\begin{align}
    D_\mathrm{KL}(r \Vert q_N) + D_\mathrm{KL}(q_N \Vert r) & {} \leqslant \varepsilon_N^2.
\end{align}
Thus the distributional deviation between reweight and target
is bounded by the quality of the degree-$N$ polynomial approximation
to $\ell$.
For Chebyshev polynomials, a standard approximation theory result
\cite[Theorem~8.3]{Trefethen:2019}
states that, if $\ell$ extends analytically to a neighbourhood of the Bernstein ellipse $E_\rho$,
$\rho > 1$,
$\varepsilon_N = \mathcal{O}(\rho^{-N})$,
and so
\begin{align}
    d_{\mathrm{TV}}(r, q_N) & {} = \mathcal{O}(\rho^{-N}),
    &
     D_\mathrm{KL}(r \Vert q_N) & {} = \mathcal{O}(\rho^{-2N}),
\end{align}
i.e.\ converge geometrically.%

The analyticity properties of $\ell$ required here are not unreasonable, since common
kinematic, resonant, etc. factors between the prior and target cancel in the ratio,
but should not be expected to hold for arbitrary IR-safe observables.
Even without geometric convergence,
the bound on the deviation of MaxEnt distributions in terms of the best polynomial
approximation error persists.
This is the case for the windowed moments for $\pT^{\ell\ell}$,
for which the target is here $C^{2}$ through \cref{eq:window}, so
$\varepsilon_N$ falls as $N^{-3}$
rather than geometrically~\cite[Theorem~7.2]{Trefethen:2019}.

Note that this applies to the exact MaxEnt problem in one dimension,
without the (small)
numerical regularisation ridge used in practice.
With additional dimensions and correlations, 
$\varepsilon_N$ corresponds to the best approximation of $\ell$
achievable within the sum of the single-variable Chebyshev constraints
considered here, and cannot vanish without cross-moments where
the target distribution has correlations beyond those accessible
within the considered class (see \textit{Cross moments} below).

\emph{Leptons and showers.}
Leptons are taken before QED
final-state radiation, the convention of the published data and of the
QCD-only fixed-order input.
POWHEG and MiNNLO$_\mathrm{PS}$ events are
showered with POWHEG emission as starting scale,
and MC@NLO with the settings Pythia recommends for 
\texttt{MG5\_aMC}. 

\begin{figure}[tb]
\includegraphics[width=\columnwidth]{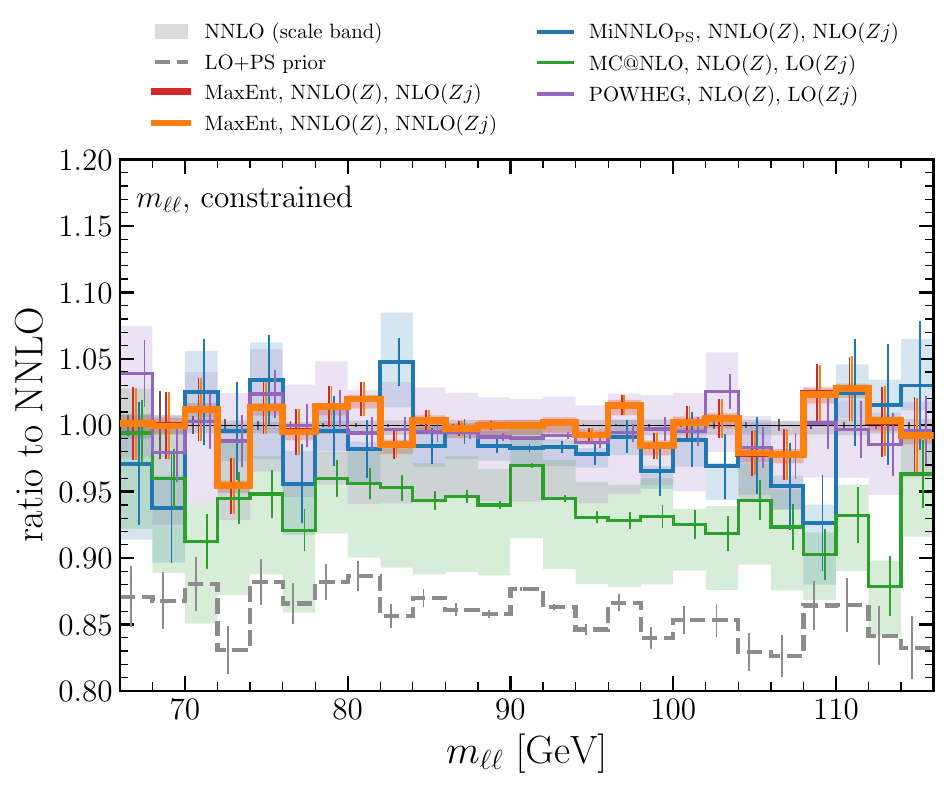}
\caption{Ratio to NNLO of the dilepton mass, constrained, with the curves, bands
and bars of \cref{fig:closure}.}
\label{fig:mll}
\end{figure}

\emph{Uncertainty propagation.}
The scale uncertainty band plotted for the MaxEnt
sample is the envelope of 
six re-solves at the standard seven-point scale variations of the
target moments, each warm-started from the central multipliers.
On the $m_{\ell\ell}$ distribution (\cref{fig:mll}),
the band closely follows that of the NNLO calculation.
Error bars add two uncertainties in quadrature: the spread of $20$ re-solves,
with targets rebuilt from the fixed-order runs (resampled with replacement),
and the reweighted sample's per-bin Monte Carlo error.
For the NNLO $Z$+jet sample the re-solves also redraw the $Z$+jet rate within its Monte Carlo error.
The $Z$+jet shape moments keep their central values, so their statistical errors do not enter the error bars.

\begin{figure}[tb]
\centering
\includegraphics[width=\columnwidth]{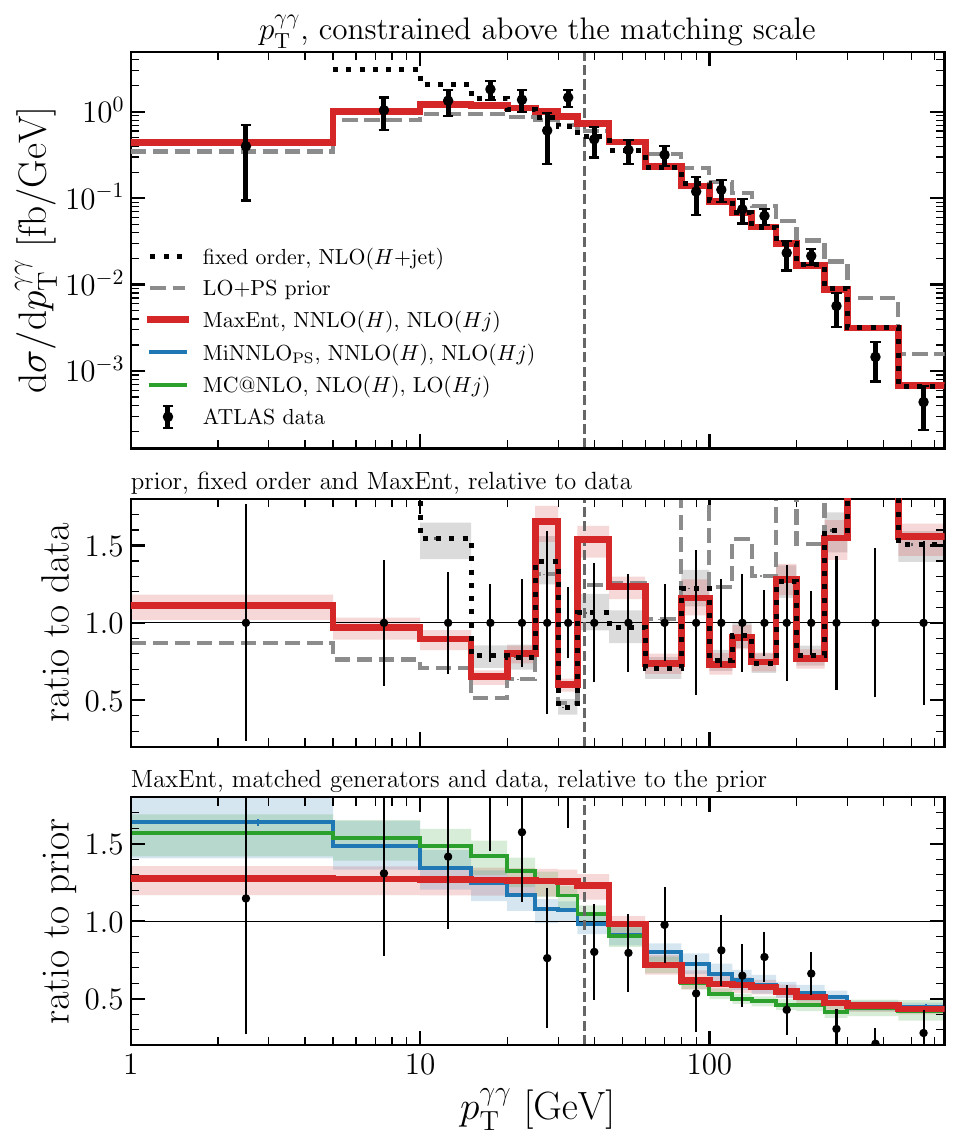}
\caption{Gluon-fusion Higgs production: the diphoton
transverse momentum in the ATLAS fiducial region~\cite{ATLAS:2022fnp},
with the curves, bands and bars of \cref{fig:money}.
These samples have been normalised to the inclusive N$^3$LO cross section~\cite{LHCHiggsCrossSectionWorkingGroup:2016ypw},
whose uncertainty is included in the bands.
}
\label{fig:higgs}
\end{figure}

\emph{Cross moments.} The construction accepts joint constraints from cross-moments
$w\,T_n(u_v)\,T_{n'}(u_{v'})$.
These carry information about the correlations
between observables that is missing from one-dimensional moment targets.
Unlike multiply-differential histogram binning,
cross-moments can resolve correlations without requiring statistically-independent integrals to converge in each bin.
For Drell--Yan we book the first four orders of $\mll$ and $|\cos\theta^{*}|$
jointly, $16$ cross-moments. On average they lie $4.1\,\sigma_\mathrm{FO}$ from fixed order
in the prior and $2.0\,\sigma_\mathrm{FO}$ after the tilt on the marginals alone, with
$\sigma_\mathrm{FO}$ as in \cref{eq:SNR}. Imposing all $16$ reproduces them at no cost in
$\effN$. 

\emph{Higgs boson production.}
We further apply the MaxEnt method to gluon-fusion Higgs production with decay to two photons, again with LO+PS prior.
Here we impose the NNLO Higgs rapidity and, above a matching scale, the NLO $H$+jet recoil from NNLOJET~\cite{NNLOJET:2024}.
Each gluon-fusion sample is normalised to the inclusive N$^3$LO cross
section~\cite{LHCHiggsCrossSectionWorkingGroup:2016ypw}
for comparison with the data.
Other production modes are then added from LO+PS samples.
\Cref{fig:higgs} compares the diphoton transverse momentum with ATLAS
data~\cite{ATLAS:2022fnp}
and 
MiNNLO$_\mathrm{PS}$~\cite{Monni:2019whf}
and
MC@NLO~\cite{Frixione:2002ik,Alwall:2014hca}
predictions,
which for the displayed distribution have NLO and LO accuracy respectively.
The MaxEnt tilts $q_i/p_i$ lie between $0.27$ and $2.0$, with $\effN/N=88\%$. MiNNLO$_\mathrm{PS}$ and MC@NLO have negative-weight fractions $f_-=14\%$ and $9\%$ and keep $\effN/N=8\%$ and $69\%$.

\bibliography{refs}

@article{Frixione:2002ik,
    author = "Frixione, Stefano and Webber, Bryan R.",
    title = "{Matching NLO QCD computations and parton shower simulations}",
    eprint = "hep-ph/0204244",
    archivePrefix = "arXiv",
    doi = "10.1088/1126-6708/2002/06/029",
    journal = "JHEP",
    volume = "2002",
    number = "06",
    pages = "029",
    year = "2002"
}

@article{Nason:2004rx,
    author = "Nason, Paolo",
    title = "{A New method for combining NLO QCD with shower Monte Carlo algorithms}",
    eprint = "hep-ph/0409146",
    archivePrefix = "arXiv",
    doi = "10.1088/1126-6708/2004/11/040",
    journal = "JHEP",
    volume = "2004",
    number = "11",
    pages = "040",
    year = "2004"
}

@article{Frixione:2007vw,
    author = "Frixione, Stefano and Nason, Paolo and Oleari, Carlo",
    title = "{Matching NLO QCD computations with Parton Shower simulations: the POWHEG method}",
    eprint = "0709.2092",
    archivePrefix = "arXiv",
    primaryClass = "hep-ph",
    doi = "10.1088/1126-6708/2007/11/070",
    journal = "JHEP",
    volume = "2007",
    number = "11",
    pages = "070",
    year = "2007"
}

@article{Alioli:2010xd,
    author = "Alioli, Simone and Nason, Paolo and Oleari, Carlo and Re, Emanuele",
    title = "{A general framework for implementing NLO calculations in shower Monte Carlo programs: the POWHEG BOX}",
    eprint = "1002.2581",
    archivePrefix = "arXiv",
    primaryClass = "hep-ph",
    doi = "10.1007/JHEP06(2010)043",
    journal = "JHEP",
    volume = "2010",
    number = "06",
    pages = "043",
    year = "2010"
}

@article{Alwall:2014hca,
    author = "Alwall, J. and Frederix, R. and Frixione, S. and Hirschi, V. and Maltoni, F. and Mattelaer, O. and Shao, H.-S. and Stelzer, T. and Torrielli, P. and Zaro, M.",
    title = "{The automated computation of tree-level and next-to-leading order differential cross sections, and their matching to parton shower simulations}",
    eprint = "1405.0301",
    archivePrefix = "arXiv",
    primaryClass = "hep-ph",
    doi = "10.1007/JHEP07(2014)079",
    journal = "JHEP",
    volume = "2014",
    number = "07",
    pages = "079",
    year = "2014"
}

@article{Jadach:2015mza,
    author = "Jadach, S. and P{\l}aczek, W. and Sapeta, S. and Si{\'o}dmok, A. and Skrzypek, M.",
    title = "{Matching NLO QCD with parton shower in Monte Carlo scheme --- the KrkNLO method}",
    eprint = "1503.06849",
    archivePrefix = "arXiv",
    primaryClass = "hep-ph",
    doi = "10.1007/JHEP10(2015)052",
    journal = "JHEP",
    volume = "2015",
    number = "10",
    pages = "052",
    year = "2015"
}

@article{Sarmah:2024hdk,
    author = "Sarmah, Pratixan and Si{\'o}dmok, Andrzej and Whitehead, James",
    title = "{KrkNLO matching for colour-singlet processes}",
    eprint = "2409.16417",
    archivePrefix = "arXiv",
    primaryClass = "hep-ph",
    doi = "10.1007/JHEP01(2025)062",
    journal = "JHEP",
    volume = "2025",
    number = "01",
    pages = "062",
    year = "2025"
}

@article{vanBeekveld:2025lpz,
	author = "van Beekveld, Melissa and Ferrario Ravasio, Silvia and Helliwell, Jack and Karlberg, Alexander and Salam, Gavin P. and Scyboz, Ludovic and Soto-Ontoso, Alba and Soyez, Gregory and Zanoli, Silvia",
	title = "{Logarithmically-accurate and positive-definite NLO shower matching}",
	eprint = "2504.05377",
	archivePrefix = "arXiv",
	primaryClass = "hep-ph",
	reportNumber = "CERN-TH-2025-004, OUTP-25-01P, Nikhef 2025-003",
	doi = "10.1007/JHEP10(2025)038",
	journal = "JHEP",
    volume = "2025",
	number = "10",
	pages = "038",
	year = "2025"
}

@article{Nason:2021xke,
	author = "Nason, Paolo and Salam, Gavin P.",
	title = "{Multiplicative-accumulative matching of NLO calculations with parton showers}",
	eprint = "2111.03553",
	archivePrefix = "arXiv",
	primaryClass = "hep-ph",
	reportNumber = "OUTP-21-26P",
	doi = "10.1007/JHEP01(2022)067",
	journal = "JHEP",
    volume = "2022",
	number = "01",
	pages = "067",
	year = "2022"
}

@article{Che:2026mep,
	author = "Che, Yuxiao and Frederix, Rikkert",
	title = "{MAcNLOPS for ZZ pair production at the LHC}",
	eprint = "2605.22108",
	archivePrefix = "arXiv",
	primaryClass = "hep-ph",
	doi = "10.1140/epjc/s10052-026-16152-7",
	journal = "Eur. Phys. J. C",
	volume = "86",
	number = "7",
	pages = "870",
	year = "2026"
}

@article{Frederix:2023hom,
	author = "Frederix, Rikkert and Torrielli, Paolo",
	title = "{A new way of reducing negative weights in MC@NLO}",
	eprint = "2310.04160",
	archivePrefix = "arXiv",
	primaryClass = "hep-ph",
	doi = "10.1140/epjc/s10052-023-12243-x",
	journal = "Eur. Phys. J. C",
	volume = "83",
	number = "11",
	pages = "1051",
	year = "2023"
}

@article{Alioli:2010qp,
	author = "Alioli, Simone and Nason, Paolo and Oleari, Carlo and Re, Emanuele",
	title = "{Vector boson plus one jet production in POWHEG}",
	eprint = "1009.5594",
	archivePrefix = "arXiv",
	primaryClass = "hep-ph",
	reportNumber = "DESY-10-158, SFB-CPP-10-85, IPPP-10-78, DCPT-10-156",
	doi = "10.1007/JHEP01(2011)095",
	journal = "JHEP",
    volume = "2011",
	number = "01",
	pages = "095",
	year = "2011"
}

@article{ElFarkh:2026wtx,
	author = "El Farkh, Saad and Frederix, Rikkert and Gouighri, Mohamed",
	title = "{MC@NLO event generation by reweighting unweighted born events}",
	eprint = "2602.18124",
	archivePrefix = "arXiv",
	primaryClass = "hep-ph",
	doi = "10.1140/epjc/s10052-026-16034-y",
	journal = "Eur. Phys. J. C",
	volume = "86",
	number = "7",
	pages = "751",
	year = "2026"
}

@article{Monni:2019whf,
    author = "Monni, Pier Francesco and Nason, Paolo and Re, Emanuele and Wiesemann, Marius and Zanderighi, Giulia",
    title = "{MiNNLO$_{\mathrm{PS}}$: a new method to match NNLO QCD to parton showers}",
    eprint = "1908.06987",
    archivePrefix = "arXiv",
    primaryClass = "hep-ph",
    doi = "10.1007/JHEP05(2020)143",
    journal = "JHEP",
    volume = "2020",
    number = "05",
    pages = "143",
    year = "2020",
    note = "[Erratum: JHEP 2022 (02), 031]"
}

@article{Monni:2020nks,
    author = "Monni, Pier Francesco and Re, Emanuele and Wiesemann, Marius",
    title = "{MiNNLO$_{\mathrm{PS}}$: optimizing $2\rightarrow 1$ hadronic processes}",
    eprint = "2006.04133",
    archivePrefix = "arXiv",
    primaryClass = "hep-ph",
    doi = "10.1140/epjc/s10052-020-08658-5",
    journal = "Eur. Phys. J. C",
    volume = "80",
    number = "11",
    pages = "1075",
    year = "2020"
}

@article{Hamilton:2013fea,
    author = "Hamilton, Keith and Nason, Paolo and Re, Emanuele and Zanderighi, Giulia",
    title = "{NNLOPS simulation of Higgs boson production}",
    eprint = "1309.0017",
    archivePrefix = "arXiv",
    primaryClass = "hep-ph",
    doi = "10.1007/JHEP10(2013)222",
    journal = "JHEP",
    volume = "2013",
    number = "10",
    pages = "222",
    year = "2013"
}

@article{Karlberg:2014qua,
    author = "Karlberg, Alexander and Re, Emanuele and Zanderighi, Giulia",
    title = "{NNLOPS accurate Drell--Yan production}",
    eprint = "1407.2940",
    archivePrefix = "arXiv",
    primaryClass = "hep-ph",
    doi = "10.1007/JHEP09(2014)134",
    journal = "JHEP",
    volume = "2014",
    number = "09",
    pages = "134",
    year = "2014"
}

@article{Alioli:2013hqa,
    author = "Alioli, Simone and Bauer, Christian W. and Berggren, Calvin and Tackmann, Frank J. and Walsh, Jonathan R. and Zuberi, Saba",
    title = "{Matching Fully Differential NNLO Calculations and Parton Showers}",
    eprint = "1311.0286",
    archivePrefix = "arXiv",
    primaryClass = "hep-ph",
    doi = "10.1007/JHEP06(2014)089",
    journal = "JHEP",
    volume = "2014",
    number = "06",
    pages = "089",
    year = "2014"
}

@article{Hoche:2014uhw,
    author = {H{\"o}che, Stefan and Li, Ye and Prestel, Stefan},
    title = "{Drell--Yan lepton pair production at NNLO QCD with parton showers}",
    eprint = "1405.3607",
    archivePrefix = "arXiv",
    primaryClass = "hep-ph",
    doi = "10.1103/PhysRevD.91.074015",
    journal = "Phys. Rev. D",
    volume = "91",
    number = "7",
    pages = "074015",
    year = "2015"
}

@article{HSFGenWG:2020,
    author = "Amoroso, Simone and others",
    collaboration = "HSF Physics Event Generator WG",
    title = "{Challenges in Monte Carlo Event Generator Software for High-Luminosity LHC}",
    eprint = "2004.13687",
    archivePrefix = "arXiv",
    primaryClass = "hep-ph",
    doi = "10.1007/s41781-021-00055-1",
    journal = "Comput. Softw. Big Sci.",
    volume = "5",
    number = "1",
    pages = "12",
    year = "2021"
}

@misc{Nason:2007vt,
    author = "Nason, Paolo",
    title = "{MINT: A Computer program for adaptive Monte Carlo integration and generation of unweighted distributions}",
    eprint = "0709.2085",
    archivePrefix = "arXiv",
    primaryClass = "hep-ph",
    year = "2007"
}

@article{Frederix:2020trv,
    author = "Frederix, R. and Frixione, S. and Prestel, S. and Torrielli, P.",
    title = "{On the reduction of negative weights in MC@NLO-type matching procedures}",
    eprint = "2002.12716",
    archivePrefix = "arXiv",
    primaryClass = "hep-ph",
    doi = "10.1007/JHEP07(2020)238",
    journal = "JHEP",
    volume = "2020",
    number = "07",
    pages = "238",
    year = "2020"
}

@article{Andersen:2020sjs,
    author = {Andersen, Jeppe R. and G{\"u}tschow, Christian and Maier, Andreas and Prestel, Stefan},
    title = "{A Positive Resampler for Monte Carlo events with negative weights}",
    eprint = "2005.09375",
    archivePrefix = "arXiv",
    primaryClass = "hep-ph",
    doi = "10.1140/epjc/s10052-020-08548-w",
    journal = "Eur. Phys. J. C",
    volume = "80",
    number = "11",
    pages = "1007",
    year = "2020"
}

@article{Nachman:2020fff,
    author = "Nachman, Benjamin and Thaler, Jesse",
    title = "{Neural resampler for Monte Carlo reweighting with preserved uncertainties}",
    eprint = "2007.11586",
    archivePrefix = "arXiv",
    primaryClass = "hep-ph",
    doi = "10.1103/PhysRevD.102.076004",
    journal = "Phys. Rev. D",
    volume = "102",
    number = "7",
    pages = "076004",
    year = "2020"
}

@article{Andersen:2021mvw,
    author = "Andersen, Jeppe R. and Maier, Andreas",
    title = "{Unbiased elimination of negative weights in Monte Carlo samples}",
    eprint = "2109.07851",
    archivePrefix = "arXiv",
    primaryClass = "hep-ph",
    doi = "10.1140/epjc/s10052-022-10372-3",
    journal = "Eur. Phys. J. C",
    volume = "82",
    number = "5",
    pages = "433",
    year = "2022"
}

@article{Andersen:2023cku,
    author = "Andersen, Jeppe R. and Maier, Andreas and Ma{\^\i}tre, Daniel",
    title = "{Efficient negative-weight elimination in large high-multiplicity Monte Carlo event samples}",
    eprint = "2303.15246",
    archivePrefix = "arXiv",
    primaryClass = "hep-ph",
    doi = "10.1140/epjc/s10052-023-11905-0",
    journal = "Eur. Phys. J. C",
    volume = "83",
    number = "9",
    pages = "835",
    year = "2023"
}

@misc{Danziger:2021xvr,
	author = {Danziger, Katharina and H{\"o}che, Stefan and Siegert, Frank},
	title = "{Reducing negative weights in Monte Carlo event generation with Sherpa}",
	eprint = "2110.15211",
	archivePrefix = "arXiv",
	primaryClass = "hep-ph",
	reportNumber = "MCNET-21-30, FERMILAB-PUB-21-536-T",
	month = "10",
	year = "2021",
}

@misc{Andersen:2026ppe,
	author = "Andersen, Jeppe R. and Cole, Ella and Maier, Andreas",
	title = "{Precision Cell Resampling with a Relative and Resonant Aware Metric}",
	eprint = "2605.13237",
	archivePrefix = "arXiv",
	primaryClass = "hep-ph",
	reportNumber = "IPPP/26/38",
	month = "5",
	year = "2026",
}

@misc{Doherty:2026fli,
	author = "Doherty, Regan and Hay, Lauren and Jain, Rishabh and LeBlanc, Matt and Marrinan, Julia and Mauceri, Camille and Roloff, Jennifer",
	title = "{Optimal-Transport-Based Cell Resampling for Negative and Pathological Event Weights}",
	eprint = "2607.08723",
	archivePrefix = "arXiv",
	primaryClass = "hep-ph",
	month = "7",
	year = "2026",
}

@article{Palmer:2025jmb,
	author = "Palmer, Christopher and Kronheim, Braden",
	title = "{Improving statistical precision in Monte~Carlo samples with negative weights via reweighting and uncertainty quantification}",
	eprint = "2510.16217",
	archivePrefix = "arXiv",
	primaryClass = "hep-ex",
	doi = "10.1103/k8w6-wn37",
	journal = "Phys. Rev. D",
	volume = "113",
	number = "1",
	pages = "012003",
	year = "2026"
}

@misc{Shyamsundar:2025nzn,
	author = "Shyamsundar, Prasanth",
	title = "{ARCANE Reweighting: A Monte Carlo Technique to Tackle the Negative Weights Problem in Collider Event Generation}",
	eprint = "2502.08052",
	archivePrefix = "arXiv",
	primaryClass = "hep-ph",
	reportNumber = "FERMILAB-PUB-25-0024-ETD",
	month = "2",
	year = "2025",
}

@article{HildebrandtSchoenberg1933,
  author  = {Hildebrandt, T. H. and Schoenberg, I. J.},
  title   = {On Linear Functional Operations and the Moment Problem
             for a Finite Interval in One or Several Dimensions},
  journal = {Annals of Mathematics},
  volume  = {34},
  number  = {2},
  pages   = {317--328},
  year    = {1933},
  doi     = {10.2307/1968205}
}

@BOOK{Schmudgen2017,
  title     = "The Moment Problem",
  author    = "Schmüdgen, Konrad",
  publisher = "Springer International Publishing",
  series    = "Graduate Texts in Mathematics",
  edition   =  1,
  month     =  nov,
  year      =  2017,
  address   = "Cham, Switzerland",
  doi       = {10.1007/978-3-319-64546-9}
}

@article{Haviland1936,
 ISSN = {00029327, 10806377},
 URL = {http://www.jstor.org/stable/2371063},
 author = {E. K. Haviland},
 journal = {American Journal of Mathematics},
 number = {1},
 pages = {164--168},
 publisher = {The Johns Hopkins University Press},
 title = "{On the Momentum Problem for Distribution Functions in More Than One Dimension. II}",
 volume = {58},
 year = {1936}
}

@article{Hausdorff1921,
  author  = {Hausdorff, Felix},
  title   = "{Summationsmethoden und Momentfolgen. I}",
  journal = {Mathematische Zeitschrift},
  year    = {1921},
  volume  = {9},
  number  = {1},
  pages   = {74--109},
  doi     = {10.1007/BF01378337},
  url     = {https://doi.org/10.1007/BF01378337},
  issn    = {1432-1823}
}

@article{Jaynes:1957zza,
    author = "Jaynes, E. T.",
    title = "{Information Theory and Statistical Mechanics}",
    doi = "10.1103/PhysRev.106.620",
    journal = "Phys. Rev.",
    number = "4",
    volume = "106",
    pages = "620--630",
    year = "1957"
}

@article{Shore:1980,
  author = "Shore, J. E. and Johnson, R. W.",
  title = "{Axiomatic derivation of the principle of maximum entropy and the principle of minimum cross-entropy}",
  journal = "IEEE Trans. Inf. Theory",
  volume = "26",
  number = "1",
  pages = "26--37",
  year = "1980",
  doi = "10.1109/TIT.1980.1056144",
  note = "[Correction: IEEE Trans. Inf. Theory 29, 942 (1983)]"
}

@article{Csiszar:1975,
  author = "Csisz{\'a}r, I.",
  title = "{$I$-divergence geometry of probability distributions and minimization problems}",
  journal = "Ann. Probab.",
  volume = "3",
  number = "1",
  pages = "146--158",
  year = "1975",
  doi = "10.1214/aop/1176996454"
}

@article{Assi:2025ibi,
    author = {Assi, Beno{\^\i}t and H{\"o}che, Stefan and Lee, Kyle and Thaler, Jesse},
    title = "{QCD Theory Meets Information Theory}",
    eprint = "2501.17219",
    archivePrefix = "arXiv",
    primaryClass = "hep-ph",
    doi = "10.1103/gf42-qzd9",
    journal = "Phys. Rev. Lett.",
    volume = "135",
    number = "13",
    pages = "131901",
    year = "2025"
}

@article{Komiske:2017aww,
    author = "Komiske, Patrick T. and Metodiev, Eric M. and Thaler, Jesse",
    title = "{Energy flow polynomials: A complete linear basis for jet substructure}",
    eprint = "1712.07124",
    archivePrefix = "arXiv",
    primaryClass = "hep-ph",
    reportNumber = "MIT-CTP-4965",
    doi = "10.1007/JHEP04(2018)013",
    journal = "JHEP",
    volume = "2018",
    number = "04",
    pages = "013",
    year = "2018"
}

@article{Assi:2026tfh,
    author = "Assi, Beno{\^\i}t and Lee, Kyle and Thaler, Jesse",
    title = "{Improving parton shower predictions via precision moments of energy flow polynomials}",
    eprint = "2604.00084",
    archivePrefix = "arXiv",
    primaryClass = "hep-ph",
    doi = "10.1007/JHEP09(2026)191",
    journal = "JHEP",
    volume = "2026",
    number = "09",
    pages = "191",
    year = "2026"
}

@book{Kish:1965,
  author = "Kish, L.",
  title = "{Survey Sampling}",
  publisher = "John Wiley \& Sons",
  address = "New York",
  year = "1965"
}

@article{Sjostrand:2022,
    author = "Bierlich, Christian and others",
    title = "{A comprehensive guide to the physics and usage of PYTHIA 8.3}",
    eprint = "2203.11601",
    archivePrefix = "arXiv",
    primaryClass = "hep-ph",
    doi = "10.21468/SciPostPhysCodeb.8",
    journal = "SciPost Phys. Codeb.",
    volume = "2022",
    pages = "8",
    year = "2022"
}

@article{Hou:2019efy,
    author = "Hou, Tie-Jiun and others",
    title = "{New CTEQ global analysis of quantum chromodynamics with high-precision data from the LHC}",
    eprint = "1912.10053",
    archivePrefix = "arXiv",
    primaryClass = "hep-ph",
    doi = "10.1103/PhysRevD.103.014013",
    journal = "Phys. Rev. D",
    volume = "103",
    number = "1",
    pages = "014013",
    year = "2021"
}

@article{NNLOJET:2024,
    author = "Huss, Alexander and others",
    collaboration = "NNLOJET",
    title = "{NNLOJET: A parton-level event generator for jet cross sections at NNLO QCD accuracy}",
    eprint = "2503.22804",
    archivePrefix = "arXiv",
    primaryClass = "hep-ph",
    doi = "10.21468/SciPostPhysCodeb.69",
    journal = "SciPost Phys. Codeb.",
    volume = "2026",
    pages = "69",
    year = "2026"
}

@article{Gehrmann-DeRidder:2005btv,
	author = "Gehrmann-De Ridder, A. and Gehrmann, T. and Glover, E. W. Nigel",
	title = "{Antenna subtraction at NNLO}",
	eprint = "hep-ph/0505111",
	archivePrefix = "arXiv",
	reportNumber = "ZU-TH-07-05, IPPP-05-18",
	doi = "10.1088/1126-6708/2005/09/056",
	journal = "JHEP",
    volume = "2005",
	number = "09",
	pages = "056",
	year = "2005"
}

@article{Daleo:2006xa,
	author = "Daleo, A. and Gehrmann, T. and Maître, D.",
	title = "{Antenna subtraction with hadronic initial states}",
	eprint = "hep-ph/0612257",
	archivePrefix = "arXiv",
	reportNumber = "ZU-TH-24-06",
	doi = "10.1088/1126-6708/2007/04/016",
	journal = "JHEP",
    volume = "2007",
	number = "04",
	pages = "016",
	year = "2007"
}

@article{Currie:2013vh,
	author = "Currie, James and Glover, E. W. N. and Wells, Steven",
	title = "{Infrared Structure at NNLO Using Antenna Subtraction}",
	eprint = "1301.4693",
	archivePrefix = "arXiv",
	primaryClass = "hep-ph",
	reportNumber = "IPPP-12-82, ZU-TH-26-12",
	doi = "10.1007/JHEP04(2013)066",
	journal = "JHEP",
    volume = "2013",
	number = "04",
	pages = "066",
	year = "2013"
}

@article{Czakon:2010td,
    author = "Czakon, M.",
    title = "{A novel subtraction scheme for double-real radiation at NNLO}",
    eprint = "1005.0274",
    archivePrefix = "arXiv",
    primaryClass = "hep-ph",
    doi = "10.1016/j.physletb.2010.08.036",
    journal = "Phys. Lett. B",
    volume = "693",
    pages = "259--268",
    year = "2010"
}

@article{Czakon:2014oma,
    author = "Czakon, M. and Heymes, D.",
    title = "{Four-dimensional formulation of the sector-improved residue subtraction scheme}",
    eprint = "1408.2500",
    archivePrefix = "arXiv",
    primaryClass = "hep-ph",
    doi = "10.1016/j.nuclphysb.2014.11.006",
    journal = "Nucl. Phys. B",
    volume = "890",
    pages = "152--227",
    year = "2014"
}

@article{Czakon:2019tmo,
    author = "Czakon, Micha{\l} and van Hameren, Andreas and Mitov, Alexander and Poncelet, Rene",
    title = "{Single-jet inclusive rates with exact color at $\mathcal{O}(\alpha_s^4)$}",
    eprint = "1907.12911",
    archivePrefix = "arXiv",
    primaryClass = "hep-ph",
    doi = "10.1007/JHEP10(2019)262",
    journal = "JHEP",
    volume = "2019",
    number = "10",
    pages = "262",
    year = "2019"
}

@article{ATLAS:2019zci,
    author = "Aad, Georges and others",
    collaboration = "ATLAS",
    title = "{Measurement of the transverse momentum distribution of Drell--Yan lepton pairs in proton--proton collisions at $\sqrt{s}=13$ TeV with the ATLAS detector}",
    eprint = "1912.02844",
    archivePrefix = "arXiv",
    primaryClass = "hep-ex",
    doi = "10.1140/epjc/s10052-020-8001-z",
    journal = "Eur. Phys. J. C",
    volume = "80",
    number = "7",
    pages = "616",
    year = "2020"
}

@article{Banfi:2010cf,
    author = "Banfi, A. and Redford, S. and Vesterinen, M. and Waller, P. and Wyatt, T. R.",
    title = "{Optimisation of variables for studying dilepton transverse momentum distributions at hadron colliders}",
    eprint = "1009.1580",
    archivePrefix = "arXiv",
    primaryClass = "hep-ex",
    doi = "10.1140/epjc/s10052-011-1600-y",
    journal = "Eur. Phys. J. C",
    volume = "71",
    pages = "1600",
    year = "2011"
}

@article{Collins:1977iv,
    author = "Collins, John C. and Soper, Davison E.",
    title = "{Angular Distribution of Dileptons in High-Energy Hadron Collisions}",
    doi = "10.1103/PhysRevD.16.2219",
    journal = "Phys. Rev. D",
    volume = "16",
    pages = "2219",
    year = "1977"
}

@article{Dasgupta:2020fwr,
	author = "Dasgupta, Mrinal and Dreyer, Fr{\'e}d{\'e}ric A. and Hamilton, Keith and Monni, Pier Francesco and Salam, Gavin P. and Soyez, Gregory",
	title = "{Parton showers beyond leading logarithmic accuracy}",
	eprint = "2002.11114",
	archivePrefix = "arXiv",
	primaryClass = "hep-ph",
	reportNumber = "CERN-TH-2020-026",
	doi = "10.1103/PhysRevLett.125.052002",
	journal = "Phys. Rev. Lett.",
	volume = "125",
	number = "5",
	pages = "052002",
	year = "2020"
}

@article{Forshaw:2020wrq,
	author = {Forshaw, Jeffrey R. and Holguin, Jack and Pl{\"a}tzer, Simon},
	title = "{Building a consistent parton shower}",
	eprint = "2003.06400",
	archivePrefix = "arXiv",
	primaryClass = "hep-ph",
	reportNumber = "MAN/HEP/2020/002, UWTHPH-2020-8, MCnet-20-11",
	doi = "10.1007/JHEP09(2020)014",
	journal = "JHEP",
    volume = "2020",
	number = "09",
	pages = "014",
	year = "2020"
}

@article{Herren:2022jej,
	author = {Herren, Florian and H{\"o}che, Stefan and Krauss, Frank and Reichelt, Daniel and Schönherr, Marek},
	title = "{A new approach to color-coherent parton evolution}",
	eprint = "2208.06057",
	archivePrefix = "arXiv",
	primaryClass = "hep-ph",
	reportNumber = "FERMILAB-PUB-22-556-T, IPPP/22/58, MCNET-22-14",
	doi = "10.1007/JHEP10(2023)091",
	journal = "JHEP",
    volume = "2023",
	number = "10",
	pages = "091",
	year = "2023"
}

@article{vanBeekveld:2022ukn,
	author = "van Beekveld, Melissa and Ferrario Ravasio, Silvia and Hamilton, Keith and Salam, Gavin P. and Soto-Ontoso, Alba and Soyez, Gregory and Verheyen, Rob",
	title = "{PanScales showers for hadron collisions: all-order validation}",
	eprint = "2207.09467",
	archivePrefix = "arXiv",
	primaryClass = "hep-ph",
	reportNumber = "OUTP-22-10P",
	doi = "10.1007/JHEP11(2022)020",
	journal = "JHEP",
    volume = "2022",
	number = "11",
	pages = "020",
	year = "2022"
}

@article{Hoche:2024dee,
	author = {H{\"o}che, Stefan and Krauss, Frank and Reichelt, Daniel},
	title = "{alaric parton shower for hadron colliders}",
	eprint = "2404.14360",
	archivePrefix = "arXiv",
	primaryClass = "hep-ph",
	reportNumber = "FERMILAB-PUB-24-0178-T, IPPP/24/20, MCNET-24-07",
	doi = "10.1103/PhysRevD.111.094032",
	journal = "Phys. Rev. D",
	volume = "111",
	number = "9",
	pages = "094032",
	year = "2025"
}

@article{Preuss:2024vyu,
	author = "Preuss, Christian T.",
	title = "{A partitioned dipole-antenna shower with improved transverse recoil}",
	eprint = "2403.19452",
	archivePrefix = "arXiv",
	primaryClass = "hep-ph",
	doi = "10.1007/JHEP07(2024)161",
	journal = "JHEP",
    volume = "2024",
	number = "07",
	pages = "161",
	year = "2024"
}

@article{Holguin:2026trd,
	author = {Holguin, Jack and Pl{\"a}tzer, Simon and Seymour, Michael H. and Sule, Siddharth},
	title = "{Studying the infrared behaviour of improved logarithmic accuracy parton showers with Herwig}",
	eprint = "2605.02622",
	archivePrefix = "arXiv",
	primaryClass = "hep-ph",
	reportNumber = "CERN-TH-2026-101, MCNET-26-09",
	doi = "10.1007/JHEP08(2026)146",
	journal = "JHEP",
    volume = "2026",
	number = "08",
	pages = "146",
	year = "2026"
}

@article{Czakon:2023hls,
	author = "Czakon, Micha{\l} and Kassabov, Zahari and Mitov, Alexander and Poncelet, Rene and Popescu, Andrei",
	title = "{HighTEA: high energy theory event analyser}",
	eprint = "2304.05993",
	archivePrefix = "arXiv",
	primaryClass = "hep-ph",
	reportNumber = "Cavendish-HEP-23/01, P3H-23-021, TTK-23-08",
	doi = "10.1088/1361-6471/ad64ef",
	journal = "J. Phys. G",
	volume = "51",
	number = "11",
	pages = "115002",
	year = "2024"
}

@article{Bern:2013zja,
	author = {Bern, Z. and Dixon, L. J. and Febres Cordero, F. and H{\"o}che, S. and Ita, H. and Kosower, D. A. and Maître, D.},
	title = "{Ntuples for NLO Events at Hadron Colliders}",
	eprint = "1310.7439",
	archivePrefix = "arXiv",
	primaryClass = "hep-ph",
	reportNumber = "UCLA-13-TEP-108, SLAC-PUB-15739, SB-F-422-13, IPHT-T13-228, IPPP-13-86, CERN-PH-TH-2013-243",
	doi = "10.1016/j.cpc.2014.01.011",
	journal = "Comput. Phys. Commun.",
	volume = "185",
	pages = "1443--1460",
	year = "2014"
}

@article{Carli:2010rw,
	author = "Carli, Tancredi and Clements, Dan and Cooper-Sarkar, Amanda and Gwenlan, Claire and Salam, Gavin P. and Siegert, Frank and Starovoitov, Pavel and Sutton, Mark",
	title = "{A posteriori inclusion of parton density functions in NLO QCD final-state calculations at hadron colliders: The APPLGRID Project}",
	eprint = "0911.2985",
	archivePrefix = "arXiv",
	primaryClass = "hep-ph",
	doi = "10.1140/epjc/s10052-010-1255-0",
	journal = "Eur. Phys. J. C",
	volume = "66",
	pages = "503--524",
	year = "2010"
}

@article{Britzger:2022lbf,
	author = "Britzger, D. and others",
	title = "{NNLO interpolation grids for jet production at the LHC}",
	eprint = "2207.13735",
	archivePrefix = "arXiv",
	primaryClass = "hep-ph",
	reportNumber = "CERN-TH-2022-125, IPPP/22/53, MPP-2022-80, ZU-TH 34/22",
	doi = "10.1140/epjc/s10052-022-10880-2",
	journal = "Eur. Phys. J. C",
	volume = "82",
	number = "10",
	pages = "930",
	year = "2022"
}

@article{Carrazza:2020gss,
	author = "Carrazza, S. and Nocera, E. R. and Schwan, C. and Zaro, M.",
	title = "{PineAPPL: combining EW and QCD corrections for fast evaluation of LHC processes}",
	eprint = "2008.12789",
	archivePrefix = "arXiv",
	primaryClass = "hep-ph",
	doi = "10.1007/JHEP12(2020)108",
	journal = "JHEP",
    volume = "2020",
	number = "12",
	pages = "108",
	year = "2020"
}

@misc{Jezo:2026adf,
	author = "Je{\v{z}}o, Tom{\'a}{\v{s}} and Nocera, Emanuele R. and Rabemananjara, Tanjona R. and Schwan, Christopher and Sharma, Tanishq and Wissmann, Jan",
	title = "{PineAPPLv1: fast and flexible theory predictions for present and future colliders}",
	eprint = "2606.17134",
	archivePrefix = "arXiv",
	primaryClass = "hep-ph",
	reportNumber = "MSUHEP-26-004, MS-TP-26-20",
	month = "6",
	year = "2026",
}

@article{Catani:1997xc,
    author = "Catani, S. and Webber, B. R.",
    title = "{Infrared safe but infinite: Soft gluon divergences inside the physical region}",
    eprint = "hep-ph/9710333",
    archivePrefix = "arXiv",
    reportNumber = "CAVENDISH-HEP-97-10, LPTHE-ORSAY-97-39",
    doi = "10.1088/1126-6708/1997/10/005",
    journal = "JHEP",
    volume = "1997",
    number = "10",
    pages = "005",
    year = "1997"
}

@article{doi:10.1137/0801014,
author = {Borwein, J. M. and Lewis, A. S.},
title = {Convergence of Best Entropy Estimates},
journal = {SIAM Journal on Optimization},
volume = {1},
number = {2},
pages = {191-205},
year = {1991},
doi = {10.1137/0801014},
URL = {https://doi.org/10.1137/0801014}
}

@misc{Assi:2026tqz,
    author = {Assi, Beno{\^\i}t and Bothmann, Enrico and Campbell, John and G{\"u}tschow, Christian and H{\"o}che, Stefan and Ju, Wan-Li and Knobbe, Max and Sch{\"o}nherr, Marek and Thaler, Jesse and Wagman, Michael L.},
    title = "{Drell-Yan lepton pair production from low to high transverse momentum}",
    eprint = "2610.02169",
    archivePrefix = "arXiv",
    primaryClass = "hep-ph",
    reportNumber = "FERMILAB-PUB-26-0706-T, IPPP/26/68, MIT-CTP 6120, MCNET-26-28",
    month = "10",
    year = "2026"
}

@article{Kullback:1951zyt,
    author = "Kullback, S. and Leibler, R. A.",
    title = "{On Information and Sufficiency}",
    doi = "10.1214/aoms/1177729694",
    journal = "The Annals of Mathematical Statistics",
    volume = "22",
    number = "1",
    pages = "79--86",
    year = "1951"
}

@book{Trefethen:2019,
author = {Trefethen, Lloyd N.},
title = {Approximation Theory and Approximation Practice, Extended Edition},
publisher = {Society for Industrial and Applied Mathematics},
year = {2019},
doi = {10.1137/1.9781611975949},
address = {Philadelphia, PA},
URL = {https://epubs.siam.org/doi/abs/10.1137/1.9781611975949}
}

@book{Pinsker:1964,
  author    = {Pinsker, Mark S.},
  title     = {Information and Information Stability of Random Variables and Processes},
  publisher = {Holden-Day},
  address   = {San Francisco},
  year      = {1964}
}

@article{Csiszar:1967,
  author  = {Csisz{\'a}r, Imre},
  title   = {Information-type measures of difference of probability distributions and indirect observations},
  journal = {Studia Sci. Math. Hungar.},
  volume  = {2},
  pages   = {299--318},
  year    = {1967}
}

@article{ATLAS:2022fnp,
    author = "Aad, Georges and others",
    collaboration = "ATLAS",
    title = "{Measurements of the Higgs boson inclusive and differential fiducial cross-sections in the diphoton decay channel with pp collisions at $ \sqrt{s} $ = 13 TeV with the ATLAS detector}",
    eprint = "2202.00487",
    archivePrefix = "arXiv",
    primaryClass = "hep-ex",
    reportNumber = "CERN-EP-2021-227",
    doi = "10.1007/JHEP08(2022)027",
    journal = "JHEP",
    volume = "2022",
    number = "08",
    pages = "027",
    year = "2022"
}

@techreport{LHCHiggsCrossSectionWorkingGroup:2016ypw,
    author = "de Florian, D. and others",
    collaboration = "LHC Higgs Cross Section Working Group",
    title = "{Handbook of LHC Higgs Cross Sections: 4. Deciphering the Nature of the Higgs Sector}",
    eprint = "1610.07922",
    archivePrefix = "arXiv",
    primaryClass = "hep-ph",
    reportNumber = "CERN-2017-002-M, CERN-2017-002",
    institution = "CERN",
    doi = "10.23731/CYRM-2017-002",
    series = "CERN Yellow Reports: Monographs",
    volume = "2",
    year = "2017"
}

@article{Assi:2023rbu,
    author = {Assi, Beno{\^\i}t and H{\"o}che, Stefan},
    title = "{New approach to QCD final-state evolution in processes with massive partons}",
    eprint = "2307.00728",
    archivePrefix = "arXiv",
    primaryClass = "hep-ph",
    reportNumber = "FERMILAB-PUB-23-336-T, MCNET-23-09",
    doi = "10.1103/PhysRevD.109.114008",
    journal = "Phys. Rev. D",
    volume = "109",
    number = "11",
    pages = "114008",
    year = "2024"
}

\end{document}